\documentclass[12pt]{article}
\usepackage{amsmath,amssymb}
\usepackage{graphicx}
\usepackage{subfigure}
\usepackage[pdftex]{hyperref}
\DeclareGraphicsExtensions{.eps,.bmp,.wmf,.jpg,.pdf}
\numberwithin{equation}{section}
\def\be{\begin{equation}}
\def\ee{\end{equation}}

\def\bea{\begin{eqnarray}}
\def\eea{\end{eqnarray}}

\title{Age problem in holographic dark energy}
\author{L.N. Granda\thanks{ngranda@univalle.edu.co} ,\, A. Oliveros\thanks{alexogar@univalle.edu.co} \, and\ W. Cardona\thanks{wilalbca@univalle.edu.co} \\
Department of Physics, Universidad del Valle\\ A.A. 25360, Cali,
Colombia} 
\date{}
\begin{document}
\maketitle

\begin{abstract}
\noindent We study the age problem of the universe with the holographic DE model introduced in \cite{granda}, and test the model with some known old high redshift objects (OHRO). The parameters of the model have been constrained using the SNIa, CMB and BAO data set. We found that the age of the old quasar APM 08 279+5255 at $z=3.91$ can be described by the model.
\end{abstract}

\section{Introduction}
\noindent 
The astrophysical data from distant Ia supernovae observations \cite{riess}, \cite{kowalski}, cosmic microwave background anisotropy \cite{spergel}, and large scale galaxy surveys \cite{tegmark}, \cite{tegmark2}, all indicate that the current Universe is not only expanding, it is accelerating due to some kind of  negative-pressure form of matter known as dark energy (\cite{copeland},\cite{turner},\cite{sahni},\cite{padmanabhan}). The combined analysis of cosmological observations also suggests that the universe is spatially flat, and  consists of about $\sim 1/3$
of dark matter (the known baryonic and nonbaryonic dark matter), distributed in clustered structures (galaxies, clusters of galaxies, etc.) and $\sim 2/3$ of homogeneously distributed (unclustered) dark energy with negative pressure. Despite the high percentage of the dark energy component, its nature as well as its cosmological origin remain unknown at present and a wide variety of models have been proposed to explain the nature of the dark energy and the accelerated expansion (see \cite{copeland,turner,sahni,padmanabhan} for review).
Among the different models of dark energy, the holographic dark energy approach is quite interesting as it incorporates some concepts of the quantum gravity known as the holohgraphic principle (\cite{bekenstein, thooft, bousso, cohen, susskind}),which first appeared in the context of black holes \cite{thooft} and later extended by Susskind \cite{susskind} to string theory. According to the holographic principle, the entropy of a system scales not with its volume, but with its surface area. In the cosmological context, the holographic principle will set an upper bound on the entropy of the universe \cite{fischler}. In the work \cite{cohen}, it was suggested that in quantum field theory a short distance cut-off is related to a long distance cut-off (infra-red cut-off $L$) due to the limit set by black hole formation, namely, if is the quantum zero-point energy density caused by a short distance cut-off, the total energy in a region of size $L$ should not exceed the mass of a black hole of the same size, thus $L^3\rho_\Lambda\leq LM_p^2$. Applied to the dark energy issue, if we take the whole universe into account, then the vacuum energy related to this holographic principle is viewed as dark energy, usually called holographic dark energy \cite{cohen} \cite{hsu}, \cite{li}. The largest $L$ allowed is the one saturating this inequality so that we get the holographic dark energy density $\rho_\Lambda=3c^2M_p^2L^{-2}$ where $c^2$ is a numerical constant and $M_p^{-2}=8\pi G$.\\

Choosing the Hubble horizon $H^{-1}$ as the infrared cut-off, the resulting $\rho_\Lambda$ is comparable to the observational density of dark energy \cite{horava}, \cite{hsu}. However, in \cite{hsu} it was pointed out that in this case the resulting equation-of state parameter (EoS) is equal to zero, behaving as pressureless matter which cannot give accelerated expansion. The particle horizon \cite{li} also results with an EoS parameter larger than $-1/3$, which is not enough to satisfy the current observational data, but the infrared cut-off given by the future event horizon \cite{li}, yields the desired result of accelerated expansion with an EoS parameter less than $-1/3$, despite the fact that it has problems with the causality. Another holographic DE model have been considered in \cite{sergei}, \cite{sergei1}.\\ 
Based on dimensional arguments, in \cite{granda,granda2} we have proposed an infrared cut-off for the holographic density of the form $\rho\approx\alpha H^2+\beta\dot{H}$. Though the theoretical root of the holographic dark energy is still unknown, this proposal may point in the correct direction as it can describe the dynamics of the late time cosmological evolution in 
a good agreement with the astrophysical data. Another interesting fact of this model is that the resulting Hubble parameter (and hence the total density) contains a matter and radiation component \cite{granda}, which become relevant  at high redshifts in good agreement with the BBN theory, and explain the cosmic coincidence. An important fact is that this model can exhibit quintom nature without the need to introduce any exotic matter\\
In this paper we use the cosmological constraints on the holographic dark energy model \cite{granda} obtained form
the 307 SNIa data set, CMB anisotropy and BAO (baryon acoustic oscillation) observations, to evaluate the age of the three known old high redshift objects OHRO and compare with the ages estimated by observations. This kind of test is useful to impose further constraints on the present holographic DE model.
\section{Restricting the parameters of the Model}
Let us start with the main features of the holographic dark energy model. The holographic dark energy density is given by
\begin{equation}\label{eq1}
\rho_\Lambda=3\left(\alpha H^2+\beta\dot{H}\right)
\end{equation}
\noindent where $\alpha$ and $\beta$ are constants to be determined and $H=\dot{a}/a$ is the Hubble parameter. The usual Friedmann equation is
\begin{equation}\label{eq3}
H^2=\frac{1}{3}\left(\rho_m+\rho_r+\rho_\Lambda\right)
\end{equation}
where we have taken $8\pi G=1$ and $\rho_{m}$, $\rho_{r}$ terms are the contributions of non-relativistic matter and radiation, respectively.
\noindent Setting $x=\ln{a}$, and solving the Friedmann equation (\ref{eq1}) one obtains (see \cite{granda})
\begin{equation}\label{eq4}
\tilde{H}^2=\frac{2}{3\beta-2\alpha+2}\Omega_{m0}e^{-3x}
+\frac{1}{2\beta-\alpha+1}\Omega_{r0}e^{-4x}+Ce^{-2 x(\alpha-1)/\beta}
\end{equation}
where we have introduced the scaled Hubble expansion rate $\tilde{H}=H/H_0$, and $H_0$ is the present value of the Hubble parameter (for $x=0$). Here $\Omega_{m0}=\rho_{m0}/3H_0^2$ and $\Omega_{r0}=\rho_{r0}/3H_0^2$ are the current density parameters of non-relativistic matter and radiation, and $C$ is an integration constant. 
The three constants $\alpha$, $\beta$ and $C$ are related by two conditions: the restriction imposed by the flatness condition and the current ($x=0$) holographic DE equation of state. Solving this conditions with respect to one of the parameters ($\beta$) conduce to the relations (see \cite{granda} for details).
\begin{equation}\label{eq5}
\alpha=\frac{1}{2}\left[2(1-\Omega_{m0}-\Omega_{r0})+\beta(\Omega_{r0}+3\omega_0(1-\Omega_{m0}-\Omega_{r0})+3)\right]
\end{equation}
and
\begin{equation}\label{eq6}
\begin{aligned}
C=&1-\frac{2\Omega_{m0}}{2(\Omega_{m0}+\Omega_{r0})-\beta\left[\Omega_{r0}+3\omega_0(1-\Omega_{m0}-\Omega_{r0})\right]}\\
&-\frac{2\Omega_{r0}}{2(\Omega_{m0}+\Omega_{r0})-\beta\left[\Omega_{r0}+3\omega_0(1-\Omega_{m0}-\Omega_{r0})-1\right]}
\end{aligned}
\end{equation}
Replacing this expressions for $\alpha$ and $C$ in (\ref{eq4}), and considering in what follows $\Omega_{r0}=0$, we obtain a ($\beta$, $\Omega_{m0}$, $\omega_0$)-dependent Hubble parameter, where $\omega_0$ is the present dark energy EOs parameter.\\
Next we constraint the parameter $\beta$ using the latest observational data including the joint analysis of the 307 super nova SNIa data from the union compilation set \cite{kowalski}, the cosmic microwave background (CMB) anisotropy and the baryon acoustic oscillation (BAO) observations \cite{tegmark}. To constraint $\beta$, we assume priors on the dark matter density parameter $\Omega_{m0}$ and the dark energy EOS parameter $\omega_0$, based on the well known amount of observational data which restricts this parameters with high level of confidence \cite{riess,kowalski,spergel,tegmark}. To consider the constraints from the 307 SNIa union sample, let us introduce the standard useful formulas and definitions. The theoretical distance modulus is defined as
\be\label{eq7}
\mu_{th}(z_i)=5Log_{10} D_L(z_i)+\mu_0
\ee
where $\mu_0=42.38-5Log_{10}h$, $h$ is the Hubble constant $H_0$ in units of 100 km/s/Mpc and $D_L(z)=H_0 d_L(z)/c$. The luminosity distance times $H_0$ is given by
\be\label{eq8}
d_L(z)=(1+z)\int_0^z\frac{cdz'}{\tilde{H}(z',\theta)}
\ee
where $\tilde{H}(z,\theta)$ from Eq. (\ref{eq4}) in terms of $z$ is given by
\be\label{eq9}
\tilde{H}(z,\theta)=\left[\frac{2}{3\beta-2\alpha+2}\Omega_{m0}(1+z)^3
+\frac{1}{2\beta-\alpha+1}\Omega_{r0}(1+z)^4+C(1+z)^{2(\alpha-1)/\beta}\right]^{1/2}
\ee
where $\theta\equiv(\beta,\Omega_m,\omega_0)$ (after replacing $\alpha$ and $C$ from (\ref{eq5},\ref{eq6}) with $\Omega_{r0}=0$). The statistical $\chi^2$ function (which determines likelihood function of the parameters) of the model parameters for the SNIa data is given by
\be\label{eq10}
\chi^{2}_{SN}(\theta)=\sum^{307}_{i=1}\frac{\left(\mu_{obs}(z_i)-\mu_{th}(z_i)\right)^2}{\sigma_i^2}
\ee
The $\chi^2$ function can be minimized with respect to the $\mu_0$ parameter, as it is independent of the data points and the data set \cite{nesseris}. Expanding the Eq. (\ref{eq15}) with respect to $\mu_0$ yields
\be\label{eq11}
\chi^{2}_{SN}(\theta)=A(\theta)-2\mu_0B(\theta)+\mu^2_0 C
\ee
which has a minimum for $\mu_0=B(\theta)/C$, giving 
\be\label{eq12}
\chi^{2}_{SN,min}(\theta)=\tilde{\chi}^{2}_{SN}(\theta)=A(\theta)-\frac{B(\theta)^2}{C}
\ee
with 
\be\label{eq13}
\begin{aligned}
A(\theta)=&\sum^{307}_{i=1}\frac{\left(\mu_{obs}(z_i)-\mu_{th}(z_i,\mu_0=0)\right)^2}{\sigma_i^2}\,\\
&B(\theta)=\sum^{307}_{i=1}\frac{\mu_{obs}(z_i)-\mu_{th}(z_i,\mu_0=0)}{\sigma_i^2}\,\\
&C=\sum^{307}_{i=1}\frac{1}{\sigma_i^2}
\end{aligned}
\ee
The next type of observations used to constraint the model parameters are the CMB  and BAO data. We use the CMB shift parameter R defined by \cite{bond}
\be\label{eq14}
R=\Omega^{1/2}_{m0}\int^{1090}_{0}\frac{dz}{\tilde{H}(z,\theta)}
\ee
where $z=1090$ is the redshift of the recombination \cite{komatsu}. The distance parameter $\tilde{A}$ is given by \cite{einsenstein}
\be\label{eq15}
\tilde{A}=\Omega^{1/2}_{m0}\tilde{H}(z_b)^{-1/3}\left[\frac{1}{z_b}\int^{z_b}_{0}\frac{dz}{\tilde{H}(z,\theta)}\right]^{2/3}
\ee
with $z_b=0.35$.
We turn now to constraint the constant $\beta$ (for given $\Omega_{m0}$ and $\omega_0$), using the combined data of the 307 Union SN Ia, the shift parameter
$R$ of CMB and the distance parameter $\tilde{A}$ of BAO. The total $\chi^2$ is given by 
\be\label{eq16}
\chi^{2}=\tilde{\chi}^{2}_{SN}+\chi^{2}_{CMMB}+\chi^{2}_{BAO}
\ee
The best-fit model parameters are those that minimize the total $\chi^{2}$. Here $\tilde{\chi}^{2}_{SN}$ is given by \ref{eq12}, $\chi^{2}_{CMB}$  and $\chi^2_{BAO}$ are given by
\be\label{eq17}
\chi^{2}_{CMB}=\frac{(R-R_{obs})^2}{\sigma^2_R},\,\,\,\chi^{2}_{BAO}=\frac{(\tilde{A}-\tilde{A}_{obs})^2}{\sigma^2_{\tilde{A}}}
\ee
The SDSS BAO measurement (\cite{einsenstein}) gives the observed value of
$\tilde{A} = 0.469(n_s/0.98)^{-0.35} \pm 0.017$ with the spectral index $n_s$ as measured by WMAP5 \cite{komatsu}, taken to be $n_s = 0.960$. The value of the shift parameter $R$ has also been updated by WMAP5 \cite{komatsu} to be $1.710\pm 0.019$. The table I shows the best fit value for $\beta$ with $1\sigma$ uncertainty, $\alpha$ and $C$, assuming priors for $\Omega_{m0}$ and $\omega_0$.
\newpage
\begin{table}[htbp]
	\centering
		\begin{tabular}{|c|c|c|c|c|}
		\hline \hline

$\Omega_m$ & $\omega_0$ & $\beta(1\sigma)$     & $\alpha$   & C          \\ \hline \hline
$0.28$     & $ -1$      & $0.625^{+0.023}_{-0.023}$  & $0.983$ & $0.707$ \\ \hline    
$0.28$     & $-1.2$   & $0.483^{+0.020}_{-0.020}$  & $0.819$  & $0.691$ \\ \hline

$0.22$     & $-0.91$    & $0.781^{+0.031}_{-0.031}$  & $1.12$& $0.791$ \\ \hline
$0.22$     &  $-0.92 $     & $0.768^{+0.030}_{-0.030}$  & $1.05$& $0.789$ \\ \hline

$0.21$      &  $-0.91$   & $0.794^{+0.03}_{-0.03}$  & $1.125$   & $0.803$ \\ \hline
$0.21$      &  $ -0.92$    & $0.781^{+0.03}_{-0.03}$  & $1.11$ & $0.802$ \\ \hline

$0.27$			&  $-1$  &  $0.633^{+0.026}_{-0.026}$ &  $0.986$ & $0.720$ \\ \hline
$0.26$     &  $-1$  &  $0.641^{+0.027}_{-0.026}$  &  $0.990$ & $0.732$ \\ \hline \hline
		\end{tabular}
  	\caption{\it The best-fit values for $\beta$ with $1\sigma$ error, from the joint SNIa+CMB+BAO analysis, with different priors on $\Omega_{m0}$ and $\omega_0$.}
\end{table}
\section{Testing the model with the OHRO}
Here we consider the age problem in the present holographic DE model by comparing the ages of the three well known old high redshift objects (OHRO), with the ones estimated using the holographic model \ref{eq1}, with the parameters constrained by the joint analysis of the 307 SNIa+CMB+BAO observations \cite{kowalski,tegmark}. The three old high redshift objects are: the 3.5 Gyr old galaxy LBDS 53W091 at redshift $z=1.55$ \cite{dunlop}, the 4.0 Gyr old galaxy LBDS 53W069 at redshift $z=1.43$ \cite{dunlop1}, and the old quasar APM 08 279+5255 at $z= 3.91$ with an estimated age of $2.0-3.0$ Gyr \cite{hasinger}. These three OHRO have been used to test many dark energy models, and it was found that the ages of the two OHRO at $z=1.43$ and $z=1.55$ can be described by most of the models, whereas the object at $z=3.91$ can not be described by the known models of DE, including the $\Lambda$CDM model, giving rise to the age problem in the DE models. (in \cite{hao} the object was accommodated but by lowering the reduced Hubble constant $h$ to 0.56).
The age of the universe can be measured by the Hubble parameter as a function of the redshift $z$ and is given by
\be\label{eq18}
t(z)=\int^{\infty}_{z}\frac{d\acute{z}}{(1+\acute{z})H(\acute{z})}
\ee
or introducing the dimensionless age parameter
\be\label{eq19}
\tau(z)=H_0t(z)=\int^{\infty}_{z}\frac{d\acute{z}}{(1+\acute{z})\tilde{H}(\acute{z})}
\ee
At any redshift, the age of the universe should be larger than, or at least equal to the age of the old high redshift objects OHRO, namely $\tau(z)\geq\tau_{OHRO}$. To compare the model $\tau$ with the observational values $\tau_{OHRO}$ at the given redshifts, is useful to use the parameter $\eta=\tau(z)/\tau_{OHRO}$ (see \cite{hao}), which should be larger than $1$ in order to accommodate the object into the DE model. Note that from Eq. (\ref{eq8}), $\tau(z)$ is independent of the Hubble constant $H_0$. On the other hand, $\tau_{OHRO}$ is proportional to $H_0$, so we have some freedom in choosing the adecuate value for $H_0$ between the experimental bounds, in order to adjust the $\tau_{OHRO}$. Using the lower bound $h=0.64$ (corresponding to the result of Freedmann et al. \cite{freedman} $h=0.72\pm0.08$), the dimensionless age parameter of OHROs are $\tau_{OHRO}(1.55)=0.229$, $\tau_{OHRO}(1.43)=0.260$ and $\tau_{OHRO}(3.91)=0.131$, where we considered the lower age estimate for the old quasar APM 08 279+5255.\\
In terms of $z$ the scaled $\tilde{H}$ Hubble parameter is given by
\be\label{eq20}
\tilde{H}(z)=\left[\frac{2}{3\beta-2\alpha+2}\Omega_{m0}(1+z)^3
+\frac{1}{2\beta-\alpha+1}\Omega_{r0}(1+z)^4+C(1+z)^{2(\alpha-1)/\beta}\right]^{1/2}
\ee
replacing $\tilde{H}(z)$ in Eq. \ref{eq19}, we can generate the holographic model-values for the dimensionless age
parameter of the OHRO at $z=3.91, 1.43$, and $1.55$. 
In table II we show $\eta$ for the parameters model presented in table I for the best-fit $\beta$ corresponding to the joint SNIa+CMB+BAO data analysis.

\begin{table}[htbp]
	\centering
		\begin{tabular}{|c|c|c|c|c|c|}
		\hline \hline

$\Omega_{m0}$ & $\beta(1\sigma)$  & $\alpha$   & $\eta(3.91)$ & $\eta(1.43)$ & $\eta(1.55)$  \\ \hline \hline
$0.28$ & $0.625^{+0.023}_{-0.023}$  & $0.983$ & $0.868$ & $1.210$ & $1.291$\\ \hline    
$0.28$ & $0.483^{+0.020}_{-0.020}$  & $0.819$ & $0.840$ & $1.194$ & $1.273$\\ \hline

$0.22$ & $0.781^{+0.031}_{-0.031}$  & $1.12$ & $1.013$ & $1.387$ & $1.488$ \\ \hline
$0.22$ & $0.768^{+0.030}_{-0.030}$  & $1.05$ & $1.011$  & $1.389$  & $1.485$\\ \hline

$0.21$ & $0.794^{+0.03}_{-0.03}$    & $1.125$ & $1.043$ & $1.424$  & $1.526$ \\ \hline
$0.21$ & $0.781^{+0.03}_{-0.03}$    & $1.11$  & $1.042$ & $1.426$  & $1.527$\\ \hline

$0.27$ & $0.633^{+0.026}_{-0.026}$ & $0.986$  &  $1.006$  & $1.411$ & $1.507$ \\ \hline
$0.26$ & $0.641^{+0.027}_{-0.026}$ & $0.990$  &  $1.030$  & $1.442$ & $1.540$ \\ \hline \hline
			
		\end{tabular}
  	\caption{\it The ratio $\eta=\tau(z)/\tau_{ORHO}$ at $z=3.91$, 1.43 and 1.55, for the holographic model with 
  	the best-fit $\beta$ from the joint SNIa+CMB+BAO data analysis, assuming $h=0.64$ to evaluate the $\tau_{ORHO}$ for $\Omega_{m0}=0.28, 0.22, 0.21$ and $h=0.56$ for $\Omega_{m0}=0.27, 0.26$.}
\end{table}

\section{discussion}
We have tested the holographic DE model given by (\ref{eq1}) with the three known OHRO at redshifts $z=3.91$, 1.43 and 1.55. First we constrained the parameters of the model by using the joint SNIa+CMB+BAO data analysis, to find the best-fit $\beta$, and then trough Eqs. (\ref{eq4},\ref{eq5}) we calculated $\alpha$ and $C$ with different priors on $\Omega_{m0}$ and $\omega_0$. It was found that the old quasar APM 08 279+5255 at $z = 3.91$ can be accommodated in the model if we use lower values for the dark matter density parameter, but so that these values remain within the bounds established by at least the model-independent cluster estimate $\Omega_{m0}=0.3\pm 0.1$ \cite{carlberg}, as can be seen from table II for $\Omega_{m0}=0.21$ and $0.22$. However, this values can be ruled out by the WMAP3 bound $\Omega_{m0}=0.268\pm 0.018$ \cite{spergel1}, weakening the argument of the smaller $\Omega_{m0}$. By other hand, if we take the lower bound on $h$ set by Sandage {\it et al} (i.e. $h=0.56$) \cite{sandage}, we can go back to the accepted by the current observational bounds values for the $\Omega_{m0}$.  For instance, if we take $\Omega_{m0}=0.27, 0.26$ given in table II, we obtain $\eta=1.006, 1.03$ respectively, which again accommodates the old quasar APM 08 279+5255 into the holographic DE model. So the age problem in the present dark energy model can be solved by lowering the Hubble parameter, which is supported by the results of Sandage {\it et al} \cite{sandage}. If we evaluate the age of the universe with the present model using Eq. (\ref{eq19}) (with the lower limit $z=0$), then in all the cases presented in table I the age of the universe is of the order of $1/H_0$, which for the used values of $H_0$ gives an age between $14.8-17.4$ Gyr. So in this age range must be the universe if we want to accommodate the old quasar APM 08 279+5255 into the present holographic DE model. The age problem in the context of the holographic dark energy has been also considered in \cite{hao}, with the event horizon as the infrared cut-off.
\section*{Acknowledgments}
This work was supported by the Universidad del Valle.

\end{document}